\documentclass[aps,prl,twocolumn,superscriptaddress,showpacs,amssymb]{revtex4}
\usepackage{graphicx}
\usepackage{color}
\begin{document}
\title{Single-ion Kondo Scaling of the Coherent Fermi Liquid Regime in Ce$_{1-x}$La$_x$Ni$_2$Ge$_2$}
\author{Adam~P.~Pikul}
\affiliation{Institute of Low Temperature and Structure Research, Polish Academy of Sciences,
Wroc{\l}aw, Poland} \affiliation{Max Planck Institute for Chemical Physics of Solids, Dresden,
Germany}
\author{Ulrike~Stockert}
\author{Alexander~Steppke}
\affiliation{Max Planck Institute for Chemical Physics of Solids, Dresden, Germany}
\author{Tomasz Cichorek}
\affiliation{Institute of Low Temperature and Structure Research, Polish Academy of Sciences,
Wroc{\l}aw, Poland} \affiliation{Max Planck Institute for Chemical Physics of Solids, Dresden,
Germany}
\author{Stefanie Hartmann}
\author{Nubia~Caroca-Canales}
\author{Niels~Oeschler}
\author{Manuel~Brando}
\author{Christoph~Geibel}
\author{Frank~Steglich}
\affiliation{Max Planck Institute for Chemical Physics of Solids, Dresden, Germany}

\date{\today}

\begin{abstract}
Thermodynamic and transport properties of the La-diluted Kondo lattice CeNi$_2$Ge$_2$ were studied
in a wide temperature range. The Ce-rich alloys Ce$_{1-x}$La$_x$Ni$_2$Ge$_2$ were found to exhibit
distinct features of the coherent heavy Fermi liquid. At intermediate compositions ($0.7 \leqslant
x \leqslant 0.9$) non-Fermi liquid properties have been observed, followed by the local Fermi
liquid behavior in the dilute limit. The $4f$-electron contribution to the specific heat was found
to follow the predictions of the Kondo impurity model both in the local as well as coherent
regimes, with the characteristic Kondo temperature decreasing rapidly from about 30~K for the
parent compound CeNi$_2$Ge$_2$ to about 1~K in the most dilute samples. The specific heat does not
show any evidence for the emergence of a new characteristic energy scale related to the formation
of the coherent Kondo lattice.
\end{abstract}
 \maketitle


Strongly correlated $f$-electron systems, \emph{i.e.} intermetallic compounds based on lanthanoids
or actinoids, have been a subject of unflagging interest for more than three decades. The main
reason for that is a number of novel physical phenomena and extraordinary behaviors evidenced in
these materials at low temperatures, \emph{e.g.} heavy quasiparticles \cite{grewe}, magnetically
driven superconductivity \cite{monthoux}, non-Fermi-liquid behavior \cite{stewart1,stewart2} and
quantum criticality \cite{loehneysen2}. The Kondo interaction between the spins of conduction
electrons and the magnetic moments of the localized $f$ shells is a common ground for all these
phenomena \cite{hewson}.

In dilute Kondo systems, the magnetic ions are well separated from each other and randomly
distributed in the lattice. Therefore, the Kondo effect occurs independently at each $f$ site, and
physical properties of such compounds are a function of the characteristic Kondo temperature
$T_{\rm K}$ \cite{hewson}. In Kondo lattice systems, containing a dense, periodic sublattice of
magnetic ions, the interactions between the $f$ shells are no longer negligible. As a consequence,
the behavior of a Kondo lattice is more complex than that of a dilute Kondo alloy, and to describe
its properties more involved theoretical treatments are commonly required.

A novel approach to describe dense Kondo systems was proposed by Nakatsuji \emph{et al.}
\cite{nakatsuji1}. Based on scaling laws found for the La-diluted Kondo-lattice system CeCoIn$_5$,
they revealed a characteristic temperature $T^{\ast}$ that governs the inter-site coupling of the
$f$ shells in the coherent Kondo lattice. $T^{\ast}$ was found to be much different from the
concentration-independent single-ion $T_{\rm K}$, responsible for the on-site
$4f$--conduction-electron hybridization. This conclusion provided the basis for the development of
a phenomenological two-fluid model \cite{nakatsuji2}, which assumes the emergence of a collective
hybridization of the whole Kondo lattice, in addition to the individual hybridization, which takes
place at each $f$ site separately.


In this letter, we show that in another La-diluted Kondo lattice, CeNi$_2$Ge$_2$, the basic
assumption of the two-fluid model, i.e. the concentration-independence of the single-ion $T_{\rm
K}$, is not fulfilled. In particular, the specific heat of the Ce$_{1-x}$La$_x$Ni$_2$Ge$_2$ alloys
follows the predictions of the single-ion Kondo model in the local \emph{as well as} coherent
regimes, with $T_{\rm K}$ decreasing by one order of magnitude with increasing the La content. The
rapid reduction of $T_{\rm K}$ with increasing La concentration is confirmed by electrical
transport properties.


The experiments were performed on polycrystalline samples of Ce$_{1-x}$La$_{x}$Ni$_2$Ge$_2$,
synthesized by conventional arc melting, followed by high-temperature homogenization. The quality
of the samples was verified by means of X-ray powder diffraction. The transport properties of the
alloys were studied at temperatures ranging from room temperature down to 2~K at ambient
conditions, using a commercial Quantum Design PPMS. The heat-capacity measurements were extended to
70~mK, employing the semi-adiabatic method in a commercial Oxford Instruments $^3$He--$^4$He
dilution fridge with a home-made measuring setup \cite{wilhelm}.


The heavy-fermion compound CeNi$_2$Ge$_2$ is a well-known, magnetically non-ordered and
non-superconducting dense Kondo system with a Kondo temperature of about 30~K \cite{knopp}. A
number of experiments revealed that the system is very close to an antiferromagnetic quantum
critical point (see, e.g., Refs. \onlinecite{knebel2,gegenwart2}). However, as we briefly reported
in Ref.~\onlinecite{pikul}, partial substitution of cerium by larger lanthanum does not induce any
long-range magnetic order in the system, although the unit-cell volume of
Ce$_{1-x}$La$_{x}$Ni$_2$Ge$_2$ increases linearly with increasing the La content. Instead, large
and constant $\Delta C/T$ values, characteristic of a heavy Fermi liquid (FL) \cite{grewe}, are
observed in the La-doped samples with $0.05 \leqslant x \leqslant 0.40$ over more than one decade
of temperature. As will become clear below, this finding allows us to analyze the $4f$-contribution
to the specific heat $\Delta C$ of Ce$_{1-x}$La$_{x}$Ni$_2$Ge$_2$ in terms of the single-ion Kondo
model.


\begin{figure}
\includegraphics[width=0.9\columnwidth]{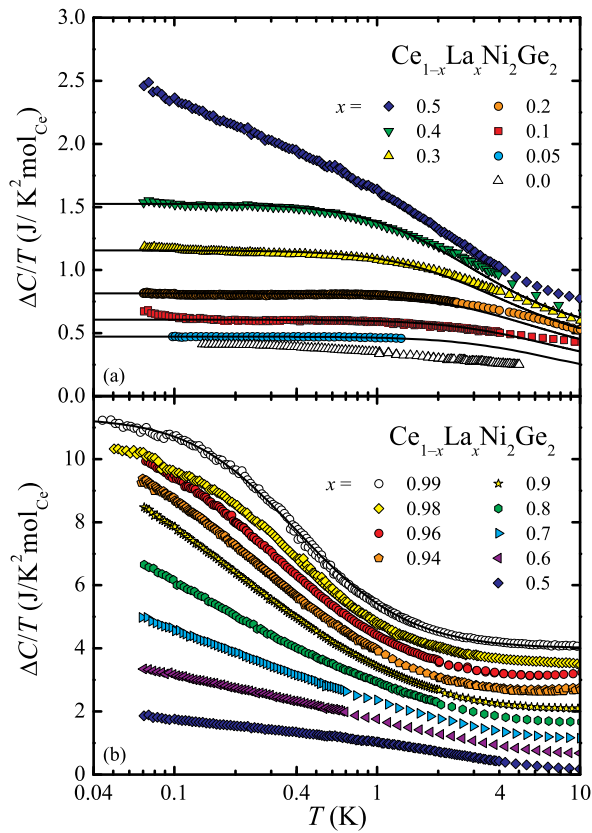}%
\caption{\label{fig1} (Color online) Temperature variation of the $4f$-electron contribution $\Delta C$ to
the total specific heat of Ce$_{1-x}$La$_x$Ni$_2$Ge$_2$ with $x\leqslant 0.50$ (a) and $x\geqslant 0.50$ (b),
normalized per mole of cerium and divided by temperature $T$. $\Delta C$ was obtained by subtraction of the
specific heat of the isostructural phonon counterpart LaNi$_2$Ge$_2$ from the raw experimental curves
(cf. Ref.~\onlinecite{pikul}). Solid lines are fits of the Kondo resonance model [Eq.~(\ref{Kondo-Cp})].
For the sake of clarity, the curves are shifted upwards by 0.1 and 0.5 J/K$^2$mol$_{\rm Ce}$ in panel (a) and (b), respectively.}
\end{figure}

\begin{figure}
\includegraphics[width=0.9\columnwidth]{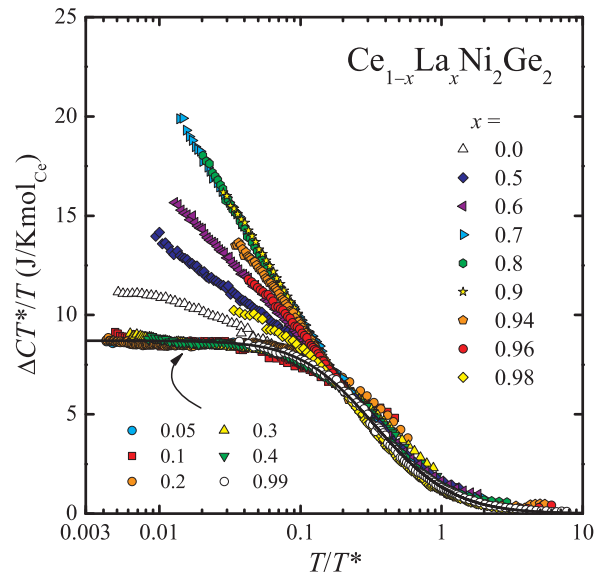}%
\caption{\label{fig2} (Color online) $\Delta C/T$ of Ce$_{1-x}$La$_{x}$Ni$_2$Ge$_2$ as a function of the normalized
temperature $T/T^{\ast}$, where $T^{\ast}$ is the characteristic temperature of the system. The solid line represents
the Kondo impurity contribution to the specific heat $\Delta C (T/T^{\ast})= C_{\rm KI} (T/T_{\rm K})$
given by Eq.~(\ref{Kondo-Cp}).}
\end{figure}


According to the Kondo resonant-level model by Schotte and Schotte~\cite{schotte}, the
Kondo-impurity contribution $C_{\rm KI}$ to the total specific heat per one mole of the impurities
with effective spin $S = 1/2$ is described by the formula:
\begin{equation}
\label{Kondo-Cp}
C_{\rm KI} \left( \frac{T}{T_{\rm K}} \right) = 2 R \frac{T_{\rm K}}{2\pi T}
\left[ 1 - \frac{T_{\rm K}}{2\pi T} \psi^{\prime} \left(\frac{1}{2} - \frac{T_{\rm K}}{2\pi T} \right) \right].
\end{equation}
$R$ is the universal gas constant, $\psi^{\prime}$ is the first derivative of the digamma function,
and $T_{\rm K}$ is the Kondo temperature defined as the width of the Lorentzian-shape Kondo
resonance at the Fermi level. As can be seen in Figs.~\ref{fig1} and \ref{fig2}, this model
describes surprisingly well our experimental data not only in the dilute limit ($x = 0.99$), as
expected, but also for the Ce-rich alloys ($0.05 \leqslant x \leqslant 0.40$). In the latter
concentration range, a slight upwards deviation from the theoretical calculations above 2~K becomes
apparent with decreasing $x$. The behavior of this additional contribution matches nicely the
predictions of Desgranges and Rasul \cite{desgranges2, desgranges3} for the contribution of higher
crystalline electric field (CEF) levels, once the Kondo scale $T_{\rm K}$ becomes larger than 1/10
of the CEF splitting $\Delta_{\rm CEF}$. Such an increase of $T_{\rm K}/\Delta_{\rm CEF}$ with
decreasing $x$ is indicated by the transport properties of Ce$_{1-x}$La$_{x}$Ni$_2$Ge$_2$
(Fig.~\ref{fig3}, see below).

$T_{\rm K}$ of the lowest-lying doublet, as obtained by a least-squares fit of Eq.~(\ref{Kondo-Cp})
to the data, was found to decrease with increasing the La content (cf. Fig.~\ref{fig4}). This is
expected for a dense Kondo system under volume expansion \cite{doniach2,lavagna}. It is worth
noting that the $C_{\rm KI} (T/T_{\rm K})$ dependence, found by applying the phenomenological
Schotte--Schotte model, is only a phenomenological approach to the density of states at the Fermi
level and a 1/2-effective spin. Although it is not the exact theory of the Kondo problem, it agrees
well with the numerical (and hence less convenient in use) solution of the $s\!-\!d$ model, based
on the \emph{on-site} Kondo interaction \cite{desgranges}. Therefore, $T_{\rm K}$ obtained from the
fits of Eq.~(\ref{Kondo-Cp}) appears to be a good approximation of the single-ion Kondo temperature
of the Ce$_{1-x}$La$_{x}$Ni$_2$Ge$_2$ alloys.


\begin{figure}
\includegraphics[width=0.9\columnwidth]{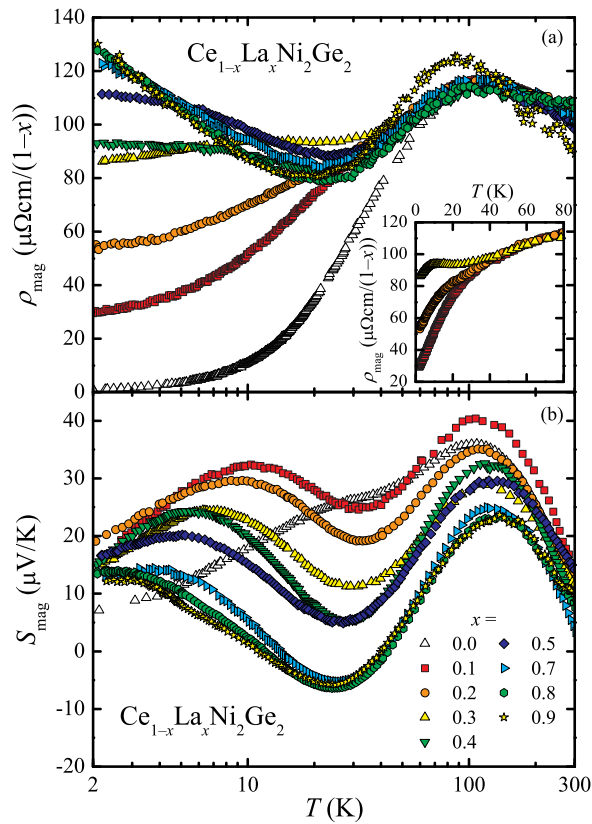}%
\caption{\label{fig3} (Color online) Temperature variations of the magnetic contribution
to the electrical resistivity (a) and thermoelectric power (b) of selected
Ce$_{1-x}$La$_{x}$Ni$_2$Ge$_2$ alloys.}
\end{figure}

\begin{figure}
\includegraphics[width=0.9\columnwidth]{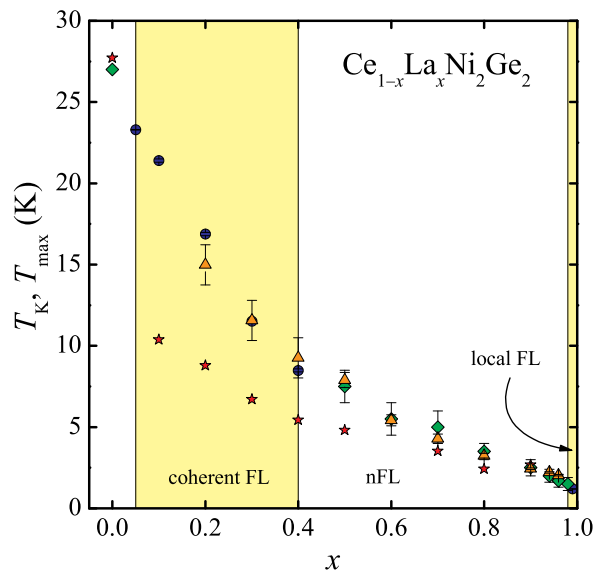}%
\caption{\label{fig4} (Color online) Tentative phase diagram of Ce$_{1-x}$La$_{x}$Ni$_2$Ge$_2$.
Circles ($\circ$) mark the single-ion $T_{\rm K}$ from the fits of the Kondo resonance model
applied to the specific-heat results both in the coherent and local FL regime (Fig.~\ref{fig1}). Diamonds ($\diamond$) are
the scaling parameter $T^{\ast}$ found for all other samples (Fig.~\ref{fig2}). Triangles
($\vartriangle$) are effective $T_{\rm K,eff}$ values estimated from entropy (Ref.~\cite{pikul})
and divided by a factor of $2\pi \! \times \! 0.103$ (cf. Ref. \cite{desgranges}). Additionally, positions
$T_{\rm max}$ of the low-temperature maxima in $S_{\rm mag}(T)$ (Fig.~\ref{fig3}) are marked by stars ($\star$).}
\end{figure}

In order to find out whether some different characteristic temperature scale potentially governs
the intermediate non-FL range, we plotted $\Delta C/T$ from Fig.~\ref{fig1} as a function of the
normalized temperature $T/T^{\ast}$, where $T^{\ast}$ is a scaling parameter (Fig.~\ref{fig2}). As
a consequence of the $T/T_{\rm K}$-dependence of $C_{\rm KI}$ [Eq.~(\ref{Kondo-Cp})], for all the
experimental curves from the two different FL regimes, a scaling relation was easily obtained with
$T^{\ast}$ being equal to $T_{\rm K}$ from the Schotte-Schotte fits. For the other curves, that do
not follow Eq.~(\ref{Kondo-Cp}), $T^{\ast}$ was chosen to get as good matching as possible. As seen
in Fig.~\ref{fig2}, these curves appeared to deviate from the theoretical predictions for
$T/T^{\ast} \lesssim 0.2$, but overlap with $C_{\rm KI} (T/T_{\rm K})$ at higher temperatures.
Moreover, for the alloys with $0.7 \leqslant x \leqslant 0.9$, the scaling relation is obeyed in
the whole temperature range studied. As inferred from Fig.~\ref{fig4}, in the non-Fermi-liquid
regime, $0.5 \leqslant x \leqslant 0.98$, the scaling displayed in Fig.~\ref{fig2} yields
$T^{\ast}$ values which smoothly interpolate between the results for the single-ion $T_{\rm K}$ in
the two adjacent FL regimes. It points out that $T_{\rm K}$ remains a dominating energy scale in
the non-FL regime.


Figure~\ref{fig3} displays the temperature dependence of the magnetic contribution to the
electrical resistivity ($\rho_{\rm mag}$) and thermoelectric power ($S_{\rm mag}$) of
Ce$_{1-x}$La$_{x}$Ni$_2$Ge$_2$. $\rho_{\rm mag}$ was calculated by subtracting the data for
LaNi$_2$Ge$_2$. $S_{\rm mag}(T)$ was determined from the Gorter-Nordheim relation $S\rho = S_{\rm
mag}\rho_{\rm mag} + S_0\rho_0$, where $S_0$ and $\rho_0$ are the data for the pure La system. At
elevated temperatures, both $\rho_{\rm mag}(T)$ and $S_{\rm mag}(T)$ exhibit broad and nearly
concentration-independent maxima just above 100~K, which can be attributed to the combined Kondo
scattering of conduction electrons off the lowest-lying and excited CEF levels. Although the exact
CEF scheme cannot be precisely determined from these data, one can roughly estimate the energy of
the first excited CEF doublet as being of the order of room temperature (cf. Ref.
\onlinecite{zlatic1}). The very weak dependence on $x$ of the high-$T$ maxima indicates that the
CEF level scheme is only weakly altered by the Ce/La substitution.

The low-temperature behavior of $\rho_{\rm mag}(T)$ and $S_{\rm mag}(T)$ of
Ce$_{1-x}$La$_{x}$Ni$_2$Ge$_2$ is in turn strongly dependent on the Ce content. In $\rho_{\rm
mag}(T)$ (Fig.~\ref{fig3}(a)), the samples with large Ce concentration exhibit some broad excess
below about 30~K, see inset of Fig.~\ref{fig3}(a). With increasing the La-content, this extra
contribution evolves into a maximum and moves towards lower temperatures. For $x=0.40$ $\rho_{\rm
mag}(T)$ saturates, and in the La-rich samples a $-\ln{T}$ slope develops in $\rho_{\rm mag}(T)$.
Such a behavior is characteristic of Kondo systems upon dilution of the magnetic sublattice (see
e.g. Ce$_{x}$La$_{1-x}$Cu$_6$ \cite{onuki}). In particular, the low-$T$ maximum observed in the
dense Ce$_{1-x}$La$_{x}$Ni$_2$Ge$_2$ alloys results from the emergence of the coherent Kondo
scattering, and the logarithmic increase in $\rho_{\rm mag}(T)$ evidenced in the diluted region
manifests the single-ion Kondo effect. The thermoelectric power of Ce$_{1-x}$La$_{x}$Ni$_2$Ge$_2$
(Fig.~\ref{fig3}(b)) exhibits a distinct low-$T$ maximum at $T_{\rm max}$ in \emph{all} the alloys
studied with $T_{\rm max}(x)$ decreasing upon increasing $x$. $S_{\rm mag}(T)$ is proportional to
$Tm$, where $m = \partial \ln{N(E)} /\partial E |_{E_{\rm F}}$ and $N(E)$ is the quasiparticle
density of states. For a dilute Kondo alloy (e.g. $x \gtrsim 0.98$), the low-$T$ $S_{\rm mag}(T)$
peak occurs at $T_{\rm max} \approx T_{\rm K}$ \cite{zlatic1}. For a Kondo lattice ($x = 0$), where
$N(E)$ develops a (partial) hybridization gap near $E_{\rm F}$ \cite{martin}, $T_{\rm max}$ refers
to the coherence temperature $T_{\rm coh}$. CeNi$_2$Ge$_2$ turns out to show $T_{\rm max}
\approx$~28~K \cite{koehler1}, almost identical to $T_{\rm K}$ (Fig.~\ref{fig4}). $T_{\rm coh}
\approx T_{\rm K}$ was also found for isostructural CeCu$_2$Si$_2$ (see e.g. Ref. \cite{stockert}).
However, the disordered Kondo-lattice system Ce$_{1-x}$La$_x$Ni$_2$Ge$_2$, $ 0.05 \lesssim x
\lesssim 0.5$, exhibits $T_{\rm max} (\approx T_{\rm coh}) < T_{\rm K}$ (Fig.~\ref{fig4}), which
reflects a more fragile coherence compared to CeNi$_2$Ge$_2$.


In conclusion, we have found that the La-doping does not induce any magnetic ordering in the
quantum-critical heavy-fermion compound CeNi$_2$Ge$_2$, although the crystal lattice expands, and
the single-ion Kondo temperature of the system rapidly decreases. Instead, the non-FL effects in
CeNi$_2$Ge$_2$ are immediately replaced, upon doping with 5\% La, by coherent FL behavior. This is
indeed very surprising, as in the Pd- and Cu-doped CeNi$_2$Ge$_2$ \cite{knebel2,buttgen} the non-FL
features occur clearly as a precursor of the antiferromagnetic order, in accordance with the
predictions of the Doniach phase diagram \cite{doniach2}. In this context, it is worth referring to
studies of two other La-doped heavy-fermion compounds, namely CeRu$_2$Si$_2$ and CeCu$_6$ (cf.
Ref.~\cite{grewe}. In the former one, slight La-doping (7\%) induces long range antiferromagnetic
order \cite{quezel}, while the latter compound remains magnetically non-ordered in a wide
concentration range \cite{onuki}. As shown by Rossat-Mignot \emph{et al.}~\cite{rossat-mignod}, the
different responses of these two systems on the La-doping are caused by different magnitudes of the
inter-site correlations, which are much stronger in CeRu$_2$Si$_2$ than in CeCu$_6$. Similar to
La-doped CeCu$_6$, the inter-site correlations in Ce$_{1-x}$La$_x$Ni$_2$Ge$_2$ become weak so
quickly that long-range magnetic order cannot form upon doping with La.

The values of $T_{\rm K}$ obtained from our specific heat study reveal a strong increase of the
single-ion Kondo scale upon decreasing average unit-cell volume. This corroborates previous studies
on (La$_{1-z}$Y$_z$)$_{1-x}$Ce$_x$Al$_2$ alloys with moderate, fixed Ce-concentration ($x$~=~0.15
and 0.06), which demonstrated a $T_{\rm K}$ increase by more than two orders of magnitude on going
from (La$_{1-x}$Ce$_x$)Al$_2$ to (Y$_{1-x}$Ce$_x$)Al$_2$ \cite{steglich2}.

Most interestingly, the coherent FL behavior is clearly visible in the Ce-rich alloys of
Ce$_{1-x}$La$_{x}$Ni$_2$Ge$_2$ over a wide $x$-range, and can be described using the single-ion
Kondo temperature as the local FL regime. The coherent Kondo scattering is inferred from our
transport results but not reflected in the temperature dependence of the specific heat. The two FL
regimes are found to be well separated by a non-FL region ($0.5 \leqslant x \leqslant 0.98$).
Whether the non-FL behavior is precursive to a low-lying magnetic phase transition, remains an open
question.

We thank S. Kirchner and S. Wirth for helpful conversations. APP acknowledges financial support of
the Alexander von Humboldt Foundation. Research in Wroc{\l}aw was supported by the Polish Ministry
of Science and Higher Education within Grant no. N~N202~102338. Research in Dresden was supported
in part by through the DFG Research Unit 960 "Quantum Phase Transitions".


\end{document}